%% file: main.tex
\documentclass[Afour,sageh,times]{sagej}
\usepackage{float} 
\usepackage{moreverb,url}
\usepackage{color}

\usepackage[colorlinks,bookmarksopen,bookmarksnumbered,citecolor=red,urlcolor=red]{hyperref}

\newcommand\BibTeX{{\rmfamily B\kern-.05em \textsc{i\kern-.025em b}\kern-.08em
T\kern-.1667em\lower.7ex\hbox{E}\kern-.125emX}}

\def\volumeyear{2023}

\begin{document}

\title{Fostering new Vertical and Horizontal IoT Applications with Intelligence Everywhere}

\author{Hung Cao\affilnum{1}, Monica Wachowicz\affilnum{1, 2}, Rene Richard\affilnum{3}, Ching-Hsien Hsu\affilnum{4}}

\affiliation{\affilnum{1}Analytics Everywhere Lab, University of New Brunswick, Canada\\
\affilnum{2}RMIT University, Australia\\
\affilnum{3}National Research Council, Canada\\
\affilnum{4}Asia University,Taiwan}

\corrauth{Hung Cao}

\email{hcao3@unb.ca}

\input{CONTENTS/0-Abstract.tex}


\maketitle

\input{CONTENTS/1-Introduction.tex}

\label{sec:II}
\input{CONTENTS/2-Section-II}

\label{sec:III}
\input{CONTENTS/3-Section-III}

\label{sec:IV}
\input{CONTENTS/4-Section-IV}

\label{sec:V}
\input{CONTENTS/5-Section-V}

\label{sec:VI}
\input{CONTENTS/6-Challenges}

\label{sec:conclusion}
\input{CONTENTS/7-Conclusion}

\begin{acks}
This research work was supported by the  NSERC/Cisco  Industrial Research  Chair,  Grant  IRCPJ  488403-1, and was partially supported by the NBIF Talent Recruitment Fund, Grant TRF 2003-001.  We would like to thank the Director of Cisco System Canada, Robert Barton, the CTO of Cisco IoT group, Russ Gyurek, and Rohan Upadhyay from Amazon Canada for the fruitful discussions on this research domain.
\end{acks}

\bibliographystyle{SageH}
\bibliography{ref.bib}

\end{document}

%% file: CONTENTS/0-Abstract.tex
\begin{abstract}
Intelligence Everywhere is predicated on the seamless integration of IoT networks transporting a vast amount of data streams through many computing resources across an edge-to-cloud continuum, relying on the orchestration of distributed machine learning models. The result is an interconnected and collective intelligent ecosystem where devices, systems, services, and users work together to support IoT applications. This paper discusses the state-of-the-art research and the principles of the Intelligence Everywhere framework for enhancing IoT applications in vertical sectors such as Digital Health,  Infrastructure, and Transportation/Mobility in the context of intelligent society (Society 5.0). It also introduces a novel perspective for the development of horizontal IoT applications, capable of running across various IoT networks while fostering collective intelligence across diverse sectors. Finally, this paper provides comprehensive insights into the challenges and opportunities for harnessing collective knowledge from real-time insights, leading to optimised processes and better overall collaboration across different IoT sectors.

\end{abstract}

%% file: CONTENTS/1-Introduction.tex
\section{Introduction}
INTERNET of Things (IoT) has secured its position as one of the most promising technologies supporting many smart city applications. In many IoT applications, communication technologies play a key role in transmitting large amounts of data streams from sensors to far-edge, edge, fog, and cloud computing resources. Currently, we have countless IoT applications in many vertical sectors, which include but are not limited to digital health, agriculture, retail, manufacturing/industrial, supply chain, energy, transportation/mobility, and intelligent homes. All of these sectors are utilising a certain communication standard and mainstream set of protocols in order to send and receive continuous and accumulated data streams \cite{ding2020iot}. 

The IoT networks are crucial for enabling effective machine learning (ML) models, ensuring they can handle the necessary ML tasks for any IoT application. However, many questions still remain around the support of IoT networks to ML models in order to generate collaboration and aggregation of diverse perspectives needed for collective intelligence to emerge. Typical questions are: Which type of IoT network fits the data life-cycle of an ML model for enabling the gathering of continuous versus accumulated data streams for ML models to foster collective intelligence? Can an ML model improve the QoS of both IoT applications and IoT networks? Which learning capability (e.g., incremental learning or federated learning) should we choose for new horizontal IoT applications?


Although more than 620 IoT platforms are available across academia and industry \cite{scully_2020}, most of the implemented ML tasks have been confined to optimization, prediction, and automation solutions. To the best of our knowledge, no IoT platform has previously been specifically designed to fully support an Intelligence Everywhere ecosystem that promotes continuous data-driven decision-making, automation, and learning across a continuum that extends from the edge (where data is generated) to the cloud (where data is stored and analyzed).

Therefore, the design of an Intelligence Everywhere ecosystem to foster new IoT applications remains an open research issue, posing several significant challenges. Limited knowledge still exists regarding the optimal selection of hardware, software, and communication components of IoT networks that can lead to the collaborative pooling of data, ML models, and insights, enabling devices and users to learn from each other and generate intelligent collective behavior.



In this paper, we introduce our proposed Intelligence Everywhere learning paradigm that can lead to advances in developing new IoT applications and the IoT network itself. The time is ripe for exploring learning capabilities capable of combining IoT networking that, if harnessed properly, may deliver the best of expectations over many vertical sectors in IoT applications and build up new horizontal IoT applications that allow users and IoT networks make collaborative decisions.

Towards this end, we have adopted Narayanan's definition of collective intelligence \cite{narayanan2022collective}, considering it as a decision-making approach in which intelligent and distributed IoT networks generate insights and feedback from their immediate environment and users. Together, they make collective decisions to perform tasks that lead to a common and desirable outcome.


The contributions of our work in this paper are as follows.

\begin{itemize}
    \item We explore the various specificities of many IoT networks and their interplay in the context of an Intelligence Everywhere learning paradigm. This paradigm is envisaged to preserve several data life-cycles of automated ML tasks by taking into account both learning and resource capabilities that are connected by IoT networks. 
    \item We compare federated learning and incremental learning to understand how they harness collective intelligence from distributed data sources, fostering collaborative decision-making and knowledge sharing from different IoT applications.
    
    \item We review three vertical sectors and their best-suited IoT networks for adopting Intelligence Everywhere learning paradigms in new IoT applications.
    \item As the final contribution, we present a comprehensive assessment of the challenges and opportunities that lie ahead, guiding future research directions in the field.
\end{itemize}

The remainder of this paper is organised as follows. Section II discusses the foundation of Intelligence Everywhere and our vision of its ecosystem. Section III discusses the important role of IoT networks in supporting Intelligence Everywhere learning paradigms. Section IV compares Intelligence Everywhere learning architectures. Next, we give several examples of potential IoT applications in vertical sectors and delineate the future perspective of horizontal IoT applications. Then, Section VI discusses several research challenges and opportunities from the viewpoint of Intelligence Everywhere Learning. Finally, the paper is concluded in Section VII.

%% file: CONTENTS/2-Section-II.tex
\section{Learning through Intelligence Everywhere}


Integrating machine learning in IoT networks containing millions of sensors can lead to significant advances in IoT applications and the network itself. Accumulated and continuous data streams need to be analysed as they are being transferred through various computing resources at the far-edge, edge, fog or cloud nodes of IoT networks. 

Data streams play a critical role in machine learning-driven IoT networks due to their impact on model accuracy and performance. Developing efficient intelligent models is becoming increasingly important to address future demands arising from IoT applications. This is especially critical due to the surge in large-scale and rapid data streams, which escalate model intricacy and lead to exponential growth in computational needs. \cite{gill2022ai} reiterate that next-generation computing systems for machine learning need to consider hardware accelerators and non-volatile memory to provide a good match between ML models and the underlying storage and computational resources. 


We argue that the next generation of IoT networks for machine learning may take full advantage of our proposed Intelligence Everywhere learning paradigm, which will rely on the integration and interoperability of four capabilities as described below:

\begin{itemize}
    \item \textit{Resource  Capability},  which consists of a  network of distributed far-edge, edge, fog, and cloud nodes connected to IoT sensors that can provide computation, storage, I/O, and processing power for the execution of automated ML tasks. 
    \item \textit{Analytical Capability},  which consists of ML models that are deployed for the execution of automated  ML analytical tasks. These tasks are integrated to enable devices and users to learn from each other and generate intelligent collective behavior. 
    \item \textit{Learning Capability}, specifically in the context of incremental and federated learning approaches, plays a crucial role in enabling IoT networks to operate in dynamic, privacy-sensitive, and resource-constrained settings.
    
    \item \textit{Data  Life-Cycle Capability},  which manages the changes that streamed data go through during the execution of the automated ML tasks.
\end{itemize}

\begin{figure*}[!ht]
    \centering
    \includegraphics[width=\linewidth]{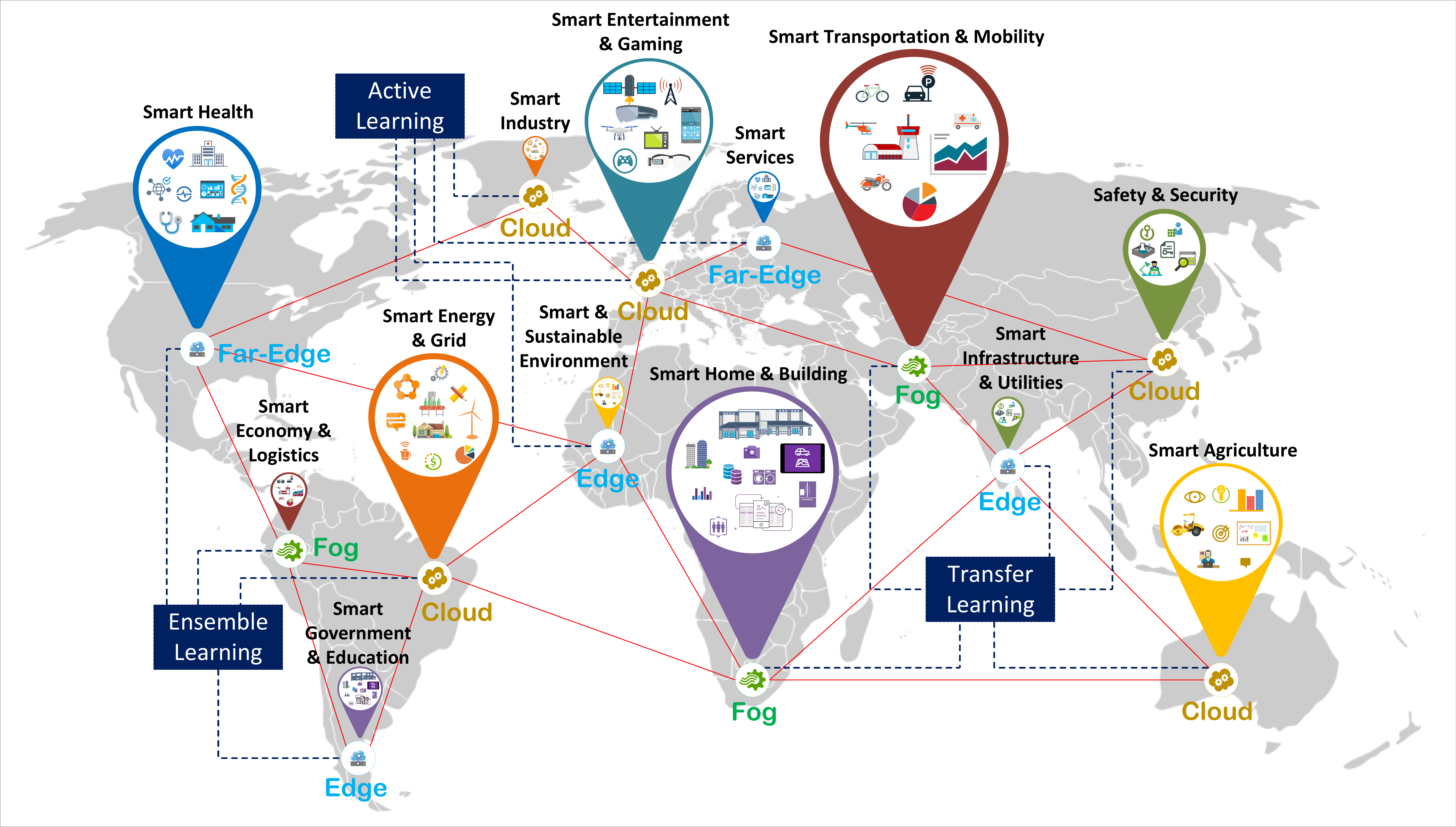}
    \caption{The vision of our proposed Intelligence Everywhere learning ecosystem.}
    \label{fig:my_label}
\end{figure*}

Fig. \ref{fig:my_label} illustrates our vision of an ecosystem where the global Intelligence Everywhere learning paradigm encompasses many platforms that are co-existing and sharing resource, analytical, learning, and data life-cycle capabilities towards an overarching goal of enabling data diversity, enhancing contextual understanding, and supporting distributed decision-making.

A sound understanding of the intrinsic capabilities of an Intelligence Everywhere ecosystem depends on
a systematic approach to gathering the massive data streams being generated from as many vertical IoT applications as possible, placing them in the context of the automated ML tasks at the right moment, and knowledgeably acting upon them. Among the latest machine learning techniques, ensemble learning has been the most researched, given its potential for dealing with the small sample sizes, high-dimensionality, and complex data structures that can be found in various IoT applications. 

Finally, it is crucial to highlight the significance of exploring how these capabilities can foster collective intelligence within our proposed dynamic learning paradigm. Towards this end, feedback loops will play a pivotal role in strengthening the collective intelligence and effectiveness of IoT networks, paving the way for the development of the next generation of IoT applications.

%% file: CONTENTS/3-Section-III.tex
\section{The important role of IoT Networks}


In this section, we consider three specificities of IoT network technologies, known as Network Architecture Design (e.g.: security protocols, associated protocols, topology, architecture), Network Performance Monitoring (e.g.: latency, QoS, bandwidth per channel, data rate, packet length), and Network Management (e.g.: power ratio, power saving mechanism).



\subsection{Network Architecture Design}

It refers to a process of planning, designing, and building an entire network. This process depends upon a set of protocols, such as associated and security protocols. Associated protocols are the ones that are working with or are supported by the respective communication technology. Security protocols are those that are solely responsible for the security of the network with respect to communication technology. For example, routing protocols for low-power and lossy networks (RPL) and advanced encryption standards (AES) are IoT-specific protocols for Wi-Fi. Depending on distinct network requirements, several types of network topologies can be used, such as point-to-point, bus, star, ring or circular, and hybrid. 

In this process, the connected resource capabilities will require always-on reachability since a heterogeneity of sensors will be connected to far-edge and edge resources using short-range networks. The data streams will be expected to stay in-memory for a limited period of time if necessary by an automated ML task, which will also depend on the data latency and the data rate of a communication network. 

Towards these challenges, \cite{guo2021enabling} suggests hyper-convergence as the best-suited network architecture for analysing real-time data streams due to its unique virtualisation and higher fault tolerance characteristics. More research is needed to study reachability as a critical requirement to return well-timed and synchronised ML tasks.  Additionally, the lack of standardization, particularly at edge resources and IoT sensors, is currently hampering the achievement of always-on reachability required for implementing the proposed Intelligence Everywhere Learning paradigm.



\subsection{Network Performance Monitoring }

Monitoring network performance is one of the key criteria in implementing our Intelligence Everywhere Learning paradigm. The scalability of these models relies on having the same network performance parameters, including latency, QoS, bandwidth per channel, data rate, compression, fragmentation capability, and packet length/capacity. The total time of latency is different for each communication technology because it is calculated by summing up three factors: (1) time taken by the data stream itself to propagate from one device to another; (2) time taken by the transmitter to transmit the content out; and (3) time taken by an end node to process the received content. QoS is another important attribute of communication technology that usually enhances the comprehensive performance of the network by giving higher preferential treatment to higher-priority traffic over the network (i.e. end-to-end). 

However, it is critical to re-think how the current practices in optimising network performances could be improved to become more suited to optimise the data life-cycle capabilities of our Intelligence Everywhere Learning ecosystem. The data life-cycles are processes for designing a seamless flow of data streams that serve as a data input and output of a sequence of distributed automated ML tasks in a network. For example, depending on the requirement of an automated ML task, we could set the priority to precise traffic like accumulated data streams. 

The majority of communication technologies support QoS features, which can be easily used to manage the performance of the automated ML tasks over the network. Data rate and packet length/capacity also play a critical role in strengthening the performance of ML tasks. Generally, packet sizes on lossy networks suffer performance issues with longer packet sizes (e.g. IPv6 1280 MTU), and communication technologies such as 6LoWPAN have endeavored to create smaller packet sizes to improve performance.

\subsection{Network Management}
Once an IoT network infrastructure is deployed and secured, network management is focused on the reliability, efficiency, and capacity of data transfer channels taking into account the protocols, applications, and computational resources of a network. For example, target wake time (TWT) is a power-saving mechanism used in Wi-Fi 6. With the help of NetFlow, a network management mechanism that tracks all the flows (i.e., a stream of packets with the same attributes, like source/destination address, port, and protocol), one can analyse the number of applications being used, and the bandwidth used by individual applications. 

In contrast, very little is known about the behaviour of the data streams during the execution of automated ML tasks. Logical specifications are needed to reflect what actually happens to both the input and output data streams when running automated ML tasks. \cite{hernandez2020uncovering} propose the use of Petri Nets to expose the actual control flow patterns behind the execution of automated ML tasks running at an edge platform. This approach provides a basis for accurate conformance checking that can enable us to foster higher confidence levels in the accuracy of executed automated ML tasks. However, the challenge still remains to differentiate when bottlenecks occur due to an inefficiency of a network performance or due to the actual execution of an automated ML task.

%% file: CONTENTS/4-Section-IV.tex
\section{Intelligence Everywhere Learning Architectures}

The key concept of Intelligence Everywhere learning is to replace the communication with data streams being sent to a cloud by a peer-to-peer communication capable of sending data streams between any edge, far-edge, and cloud resources. These data streams might contain raw data from a sensor or the input/output of an automated ML task. The network topology can be represented as a connected graph in which nodes are any computational resource and an edge indicates a communication channel between two resources. In contrast to the star graph topology, a network graph will send/receive data streams to/from a small number of nodes. 

In the context of machine learning, there is no longer a global ML model, but instead, multiple ML models that are designed to support automated ML tasks and incrementally reach a consensus at a global level \cite{cao2019developing}. This opens the opportunity to combine a variety of learning techniques, including multi-task learning, transfer learning, active learning, representation learning, online learning, and ensemble learning. It also opens the prospect of designing Intelligence Everywhere Learning paradigms that integrate descriptive, diagnostic, predictive, and prescriptive analytics by handling a variety of data sources, described as one of the following:

\begin{itemize}
    \item \textit{IoT network signaling data streams:} audio, image, and video data transmission is always accompanied by control messages, as known as signaling data. Learning about network signaling patterns will ensure the regularity, reliability, efficiency, and security of networks, as well as for automated ML tasks at the far-edge computational nodes. This is where signaling and IoT sensor data can be analysed and provide vital information to many IoT applications, such as emergency response, medical retrieval, and telecare.
    \item \textit{IoT network traffic streams:} Traffic data can be analysed from many perspectives, from that of the device model, the service provider, or the user. In general, traffic data contains several main features and characteristics, including traffic volume, downlink and uplink traffic, network access time, subscribers, flow logs, and requests. Machine learning models are useful to enhance network management, uncover diurnal patterns, and enable performance analysis and prediction, security management, and failure detection. This knowledge is paramount for ensuring the performance of automated ML tasks for IoT applications in the industrial sector.
    \item \textit{IoT localization data streams:} GPS sensors, Wi-Fi, Bluetooth signal, call details records, cell change updates, network measurement reports, and base station GIS info are examples of localization data streams. Automated ML tasks can balance network load and optimise network utilisation. It also provides support for urban planning, intelligent transportation system management, rapid emergency responses, crime prevention activities, and demographic analyses.
    \item \textit{IoT network waveforms data streams:} The 5G massive Multiple Input Multiple Output (MIMO) system is not only a communication system, but also modulated waveforms that can be analysed at the massive MIMO base stations in order to estimate the mobility of a user, whether motionless, at a slow speed, at a nearly constant speed, or at a high speed. It is tremendously useful for learning human behavior in the context of future smart cities.
    \item \textit{IoT network operational and alarm data streams:} Some peculiar communication systems, such as land mobile radio systems and RF systems, create massive amounts of system alarms and operational data daily. This is vital information for remote asset management. Analytics of this data can help to solve problems of system reliability and high maintenance costs.
\end{itemize}

It is worth noting that the Intelligence Everywhere Learning paradigm outlined in this paper still requires a learning architecture for setting up automated ML tasks that collaboratively can train an ML model running among a set of computational resources and IoT applications. Many learning mechanisms might fit into our learning paradigm, 
but there are two most feasible options to consider: federated learning \cite{mawuli2023semi} and incremental learning \cite{van2022three}, as shown in Table \ref{tab:1}.

Incremental learning allows an ML model to continuously learn from new data without forgetting the previously acquired knowledge. One of the primary advantages of utilizing incremental learning models is their ability to handle concept drift, which is crucial for maintaining accuracy over time and facilitating the generation of collective intelligent behavior in vertical IoT applications.

In contrast, federated learning is a distributed learning approach that enables multiple IoT devices or computing nodes to collaboratively train a global model while keeping their data locally. The result is a learning capability that allows the global model to learn from diverse data sources, making it adaptable to support collective intelligent behaviour in horizontal IoT applications.




\begin{table}[!ht]
\centering
\caption{Main Characteristics of Federated and Incremental Learning Architectures}
\label{tab:1}
\resizebox{\linewidth}{!}{
\begin{tabular}{lll}
\toprule
                  & Federated Learning & Incremental Learning\\\midrule
Data Distribution & \begin{tabular}[c]{@{}l@{}}Accumulated data streams\\ are gathered on multiple \\computational resources \end{tabular}                & \begin{tabular}[c]{@{}l@{}}Continuous and accumulated \\data streams are gathered on \\multiple computational resources\end{tabular}\\ \cmidrule{2-3}
Data Partitioning & Only vertical learning & horizontal and vertical learning \\\cmidrule{2-3}
Training Settings & \begin{tabular}[c]{@{}l@{}}Training a model on multiple \\computational resources. Only \\sending updates to the cloud.\end{tabular} & \begin{tabular}[c]{@{}l@{}}Training a model on multiple \\computational resources.\\ Sending updates to any other \\computational resource. \end{tabular}\\\cmidrule{2-3}
Orchestration     & Training organised in the cloud & \begin{tabular}[c]{@{}l@{}}Training organised by any \\computational resource at the \\edge, far-edge, or/and cloud \end{tabular}\\\cmidrule{2-3}
ML tasks          & \begin{tabular}[c]{@{}l@{}}Only one ML task is\\ performed (e.g. classification) \end{tabular}                                     & \begin{tabular}[c]{@{}l@{}}Different ML tasks are performed \\(e.g. classification and clustering) \end{tabular}\\\cmidrule{2-3}
Learning Process  & \begin{tabular}[c]{@{}l@{}}Centralised learning but\\ distributed training process\end{tabular}                                    & \begin{tabular}[c]{@{}l@{}}Decentralised learning and \\distributed training process\end{tabular}\\
\bottomrule
\end{tabular}}
\end{table}

It is still premature to decide which learning architecture will prevail in the future. Human-oriented design and experimental machine learning data-driven approaches are the main paths to be explored when developing new horizontal IoT applications using these learning architectures.  A holistic interdisciplinary perspective is needed to first identify which vertical sector needs have to be addressed, and then to combine these needs with communication technologies and machine learning models to amplify the potential for collective intelligence generation.

%% file: CONTENTS/5-Section-V.tex
\section{Vertical and Horizontal IoT Applications}
\label{section:Iot_app}

There are many different vertical sectors adopting IoT applications. Personal IoT applications have been deployed in smart homes by allowing owners to have the ability to control appliances or devices around the house, and wearable devices are offering personal healthcare applications for monitoring physical activities. In general, they have shown a slow adoption trend in the last decade. In contrast, government and industrial IoT applications have been widely adopted, despite requiring a collaboration among many stakeholders. The leading vertical sectors in IoT applications are manufacturing/industrial, transportation/mobility, energy, retail, digital health, supply chain, and agriculture \cite{scully_2020}. This section reviews several potential vertical sectors and their best-suited IoT networks for adopting  Intelligence Everywhere learning paradigms. They are the digital health, infrastructure, and transportation sectors.


\subsection{Digital Health Sector}
Table \ref{tab:BAN} summarises the most suitable, stable, and efficient communication technologies available for wearable-based IoT applications in the digital health sector. In a short-range communication scenario (e.g., wireless human sensing, gesture recognition), an RFID solution fits well because of its unique ID. 
With the advancement in Wi-Fi technologies, Wi-Fi 6 is a favorable solution for non-intrusive human sensing \cite{zhang2022two}. Depending on the requirements of a wearable application, various network types (e.g. BAN, WPAN, and WLAN) can be selected. Aiming to address congestion and interference issues, academic and industry researchers are working on Wi-Fi 7 technology that operates in the 2.4 GHz, 5 GHz, and 6 GHz bands with the promise of four times faster in speed compared to Wi-Fi 6 \cite{chen2022overview}. Z-wave has a -20 to 0 dBm power ratio and a sleep/awake power-saving mechanism, making it an appropriate match for wearable applications from a network management perspective. Since users could manage smart lighting from their smartwatches (part of BAN), z-wave (WPAN) would then give access to a smart hub. 
Considering energy harvesting and low cost of tags, NFC is crucial for privacy, contactless payments and other use cases such as fruit ripeness sensor, NFC sensor for pH monitoring \cite{boada2022battery}. Regarding coverage range, RF and WirepasMesh can give us the flexibility to go beyond a 100-meter range with IoT wearables. From a security perspective, 6LoWPAN is an exemplary option as it only works with IPv6.

\begin{table*}[ht!]
\centering
\caption{Communication Technologies Suited for Wearable Devices}
\label{tab:BAN}
\resizebox{\linewidth}{!}{
\begin{tabular}{llllllllll}
\toprule
\begin{tabular}[c]{@{}l@{}}\textbf{Communication} \\ \textbf{Technologies}\end{tabular} & \begin{tabular}[c]{@{}l@{}}\textbf{IPv6}\\ \textbf{Support}\end{tabular} & \textbf{Duplex Mode} & \textbf{Type of Network} & \textbf{QoS} & \textbf{Latency (ms)} & \textbf{Data Rate (Mbps)} & \begin{tabular}[c]{@{}l@{}}\textbf{Power}\\ \textbf{Ratio}\\\textbf{(dBm)}\end{tabular} & \begin{tabular}[c]{@{}l@{}}\textbf{Power} \\ \textbf{Saving} \\ \textbf{Mechanism}\end{tabular} & \textbf{Coverage Range (m)} \\
\midrule
ANT & No & Full & WPAN & No & 7.5 - 15 & 1 & \textless 17 & Yes & 30 \\
\begin{tabular}[c]{@{}l@{}}BLE (Bluetooth\\ 4/4.1/4.2/5)\end{tabular} & Yes & Half / Full & BAN / WPAN & Yes & 3, 6 & 1, 2, 3 & 0 -10 & Yes & 15 - 30, 50 - 70, 100 \\
Bluetooth & No & Half / Full & BAN / WPAN & No & 100 & 1 & 0 - 10 & Yes & 1,10,100 \\
EnOcean & Yes & Half & WPAN & No & 40 - 100 & 0.125 & 10 & Yes & 30 \\
Infrared & Yes & Half & WLAN & Yes & 175 & 12500 & 24 & No & 0.001 - 30 \\
NFC & Yes & Half & BAN / WPAN & Yes & 125 & \begin{tabular}[c]{@{}l@{}}0.106, 0.212, \\ 0.424, 0.848\end{tabular} & 20, 23 & No & 0.1 - 0.2 \\
RF & Yes & Half / Full & N/A & Yes & 4.2 & 0.02 - 10000 & 20 - 40 & No & 0.0001 - 100000000 \\
RFID & Yes & Half / Full & WPAN & Yes & 20 & 0.5 & 1.8 & Yes & 0.1 - 5 \\
Thread & Yes & Half & HAN / WPAN & No & 100 & 0.25 & 21 & Yes & 30 \\
Wi-Fi 6 & Yes & Half & LAN / CAN / WPAN / WLAN & Yes & 1.5 & $>$=1000 & 32 & Yes & 60 - 1000 \\
WirepasMesh & Yes & Half / Full & WPAN & Yes & 10 & 1 & 5 & No & 100 - 10000 \\
Zigbee & Yes & Half / Full & WPAN & Yes & 20 & 0.02, 0.04, 0.25 & 20 & Yes & 10 - 100 \\
Z-Wave & Yes & Half & WPAN & No & 3000 & 0.0096, 0.04, 0.1 & -20 - 0 & Yes & 30, 100 \\
6LoWPAN & Yes & Half & WPAN & Yes & 50 - 250 & 0.25 & 3 - 22 & Yes & 10 - 100 \\\bottomrule
\end{tabular}}
\end{table*}

\begin{figure*}[!ht]
\centering
\includegraphics[width=\linewidth]{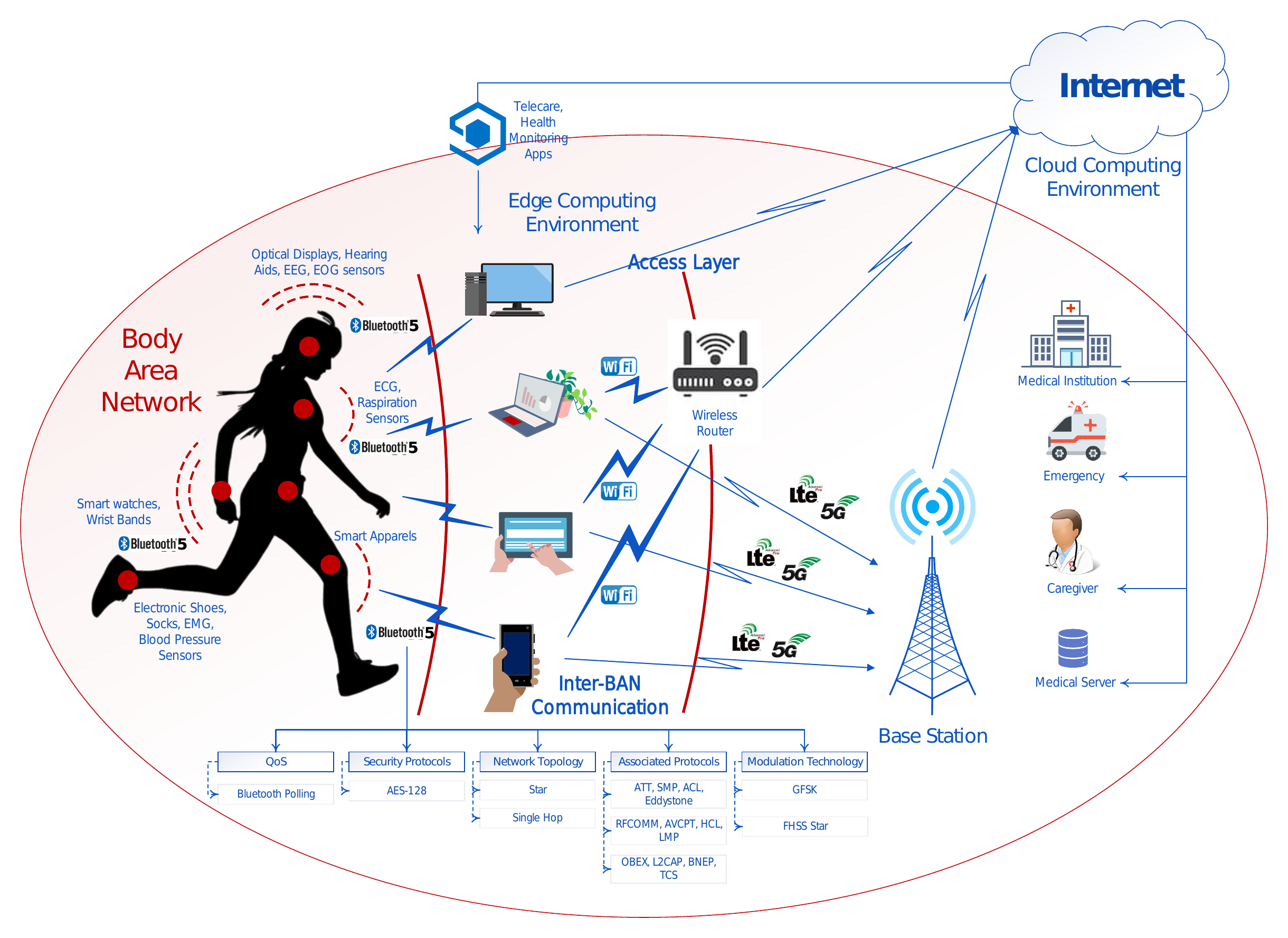}
\caption{Communication scenario for IoT wearable applications.}
\label{fig_BAN-BLE}
\end{figure*}

A typical communication scenario for IoT health applications can be found in a Body Area Network where where EoG, ECG, and EMG sensors are attached in the form of a small chipset within wearable devices like smartwatches, wrist bands, smart shoes, clothes, or earbuds. 
Fig. \ref{fig_BAN-BLE} shows a communication scenario for IoT health applications operating in BAN where EoG, ECG, and EMG sensors are attached in the form of a small chipset within wearable devices like smartwatches, wristbands, smart shoes, clothes, or earbuds. 
These gears sense, interact and transfer data to the receiving devices through Bluetooth 5 and Wi-Fi 6. Bluetooth 5 works on 2.4GHz band with half/full-duplex mode of transmission and supports IPv6 and AES-128bit security. Moreover, through polling, it provides QoS with a minimum of 3ms latency and 3Mbps of data rate within a 100 meter coverage area. Wi-Fi 6, and the integration of Wi-Fi 6 with IoT wearables, is one of the fastest-growing communication technologies because it brings kernel benefits to security, interoperability, flexibility, and improvement in data throughput. 

Most IoT applications are currently focusing on monitoring continuous data streams coming from wearable sensors for activity recognition \cite{ige2022survey}. The aim is to intuitively gather information about the behavior of users and engage them in healthy physical activities. Several ML techniques, such as one-class K-mean clustering, SVM, and CNN, are used for human/physical activity recognition. There is also a trend to analyze continuous data streams at the edge rather than in the cloud due to privacy issues \cite{cao2019analytics1}.

The adoption of wearables in IoT applications in the digital health sector has recently increased in light of the COVID-19 pandemic. We expect the demand for new IoT applications, such as wearable-enabled assistance, in response to increased care needs. Other applications, such as Aging in Place, where people will have the health and social support and services they need to live safely and independently in their homes and communities, will further highlight the importance of IoT technology. These applications will need to mimic learning that enables knowledge transfer from ML models and collective insights from federated and incremental learning mechanisms rather than sharing private wearable users’ data.

\subsection{Infrastructure Sector}

Table \ref{tab:HAN} sketches the significant communication technologies that are appropriate for medium-range coverage. Except for Wi-Fi HaLow and Wireless HART, the majority of them support the IPv6 feature. 
Several types of networks, such as WLAN, LAN, CAN, and HAN, can be considered designed and implemented in medium-range intelligent home applications \cite{du2022overview}.
Based on the data rate, latency, power consumption, and QoS specification, Z-wave, BLE, Zigbee, and Thread are the preferred technologies for the medium-range IoT system \cite{Domb2019SmartHome}. Regarding wired communication, ethernet and optical fiber are considered as a backbone for such IoT use cases. WirepasMesh and DigiMesh are both well-matched for a mesh network, although we must consider the fact that the power-saving mechanism is not supported by WirepasMesh, which is an important disadvantage.

\begin{table*}[ht!]
\centering
\caption{ Communication Technologies Suited for Intelligent Home Applications.}
\label{tab:HAN}
\resizebox{\linewidth}{!}{
\begin{tabular}{llllllllll}
\toprule
\begin{tabular}[c]{@{}l@{}}\textbf{Communication} \\ \textbf{Technologies}\end{tabular} & \textbf{IPv6 Support} & \textbf{Duplex Mode} & \textbf{Type of Network} & \textbf{QoS} & \textbf{Latency (ms)} & \textbf{Data Rate (Mbps)} & \textbf{Power Ratio (dBm)} & \begin{tabular}[c]{@{}l@{}}\textbf{Power} \\ \textbf{Saving} \\ \textbf{Mechanism}\end{tabular} & \textbf{Coverage Range (m)} \\
\midrule
\begin{tabular}[c]{@{}l@{}}BLE (Bluetooth\\ 4/4.1/4.2/5)\end{tabular} & Yes & Half / Full & BAN / WPAN & Yes & 3, 6 & 1, 2, 3 & 0-10 & Yes & 15 - 30, 50 - 70, 100 \\
Bluetooth & No & Half / Full & BAN / WPAN & No & 100 & 1 & 0 - 10 & Yes & 1,10,100 \\
DigiMesh & Yes & Half / Full & WPAN & No & 13 & 0.1152 & 10 - 18 & Yes & 90 - 1600 \\
Ethernet & Yes & Half / Full & LAN / MAN & Yes & 5 - 125 & 1 - 400000 & 40 - 48 & No & 100 \\
Optical Fiber & Yes & Half / Full & LAN / MAN / WAN & Yes & 0.005 & 100 - 43000000 & 30 - 37 & Yes & 550, 1000, 2000, 10000 \\
RF & Yes & Half / Full & N/A & Yes & 4.2 & 0.02 - 10000 & 20 - 40 & No & 0.0001 - 100000000 \\
Thread & Yes & Half & HAN / WPAN & No & 100 & 0.25 & 21 & Yes & 30 \\
Wi-Fi 6 & Yes & Half & LAN / CAN / WPAN / WLAN & Yes & 1.5 & $>$=1000 & 32 & Yes & 60 - 1000 \\
Wi-Fi  & Yes & Half & LAN / CAN / WPAN / WLAN & Yes & \begin{tabular}[c]{@{}l@{}}1.58, 7.89, 34.87, \\ 58.91, 190 \end{tabular} & \begin{tabular}[c]{@{}l@{}}a/g: 54, b:11, n:600, \\ac:1000, ad:7000, af:568.9, \\ah:347, ax:10000, ay:100000 \end{tabular} & 32 & Yes & 10 - 1000 \\
WirepasMesh & Yes & Half / Full & WPAN & Yes & 10 & 1 & 5 & No & 100 - 10000 \\
Wi-Fi HLow & No & Half & CAN / WLAN & Yes & 64 & 0.15 – 4, 0.65 - 7.8 & $<$30 & Yes & 100 – 1000, 700 \\
Wireless HART & No & Half & CAN / WLAN & Yes & 500 - 1000 & 0.25 & 10 & Yes & 10 - 600 \\
Zigbee & Yes & Half / Full & WPAN & Yes & 20 & 0.02, 0.04, 0.25 & 20 & Yes & 10 - 100 \\
Z-Wave & Yes & Half & WPAN & No & 3000 & 0.0096, 0.04, 0.1 & 0 & Yes & 30, 100 \\
6LoWPAN & Yes & Half & WPAN & Yes & 50 - 250 & 0.25 & 3 - 22 & Yes & 10 - 100 \\\bottomrule
\end{tabular}}
\end{table*}

\begin{figure*}[!ht]
\centering
\includegraphics[width=\linewidth]{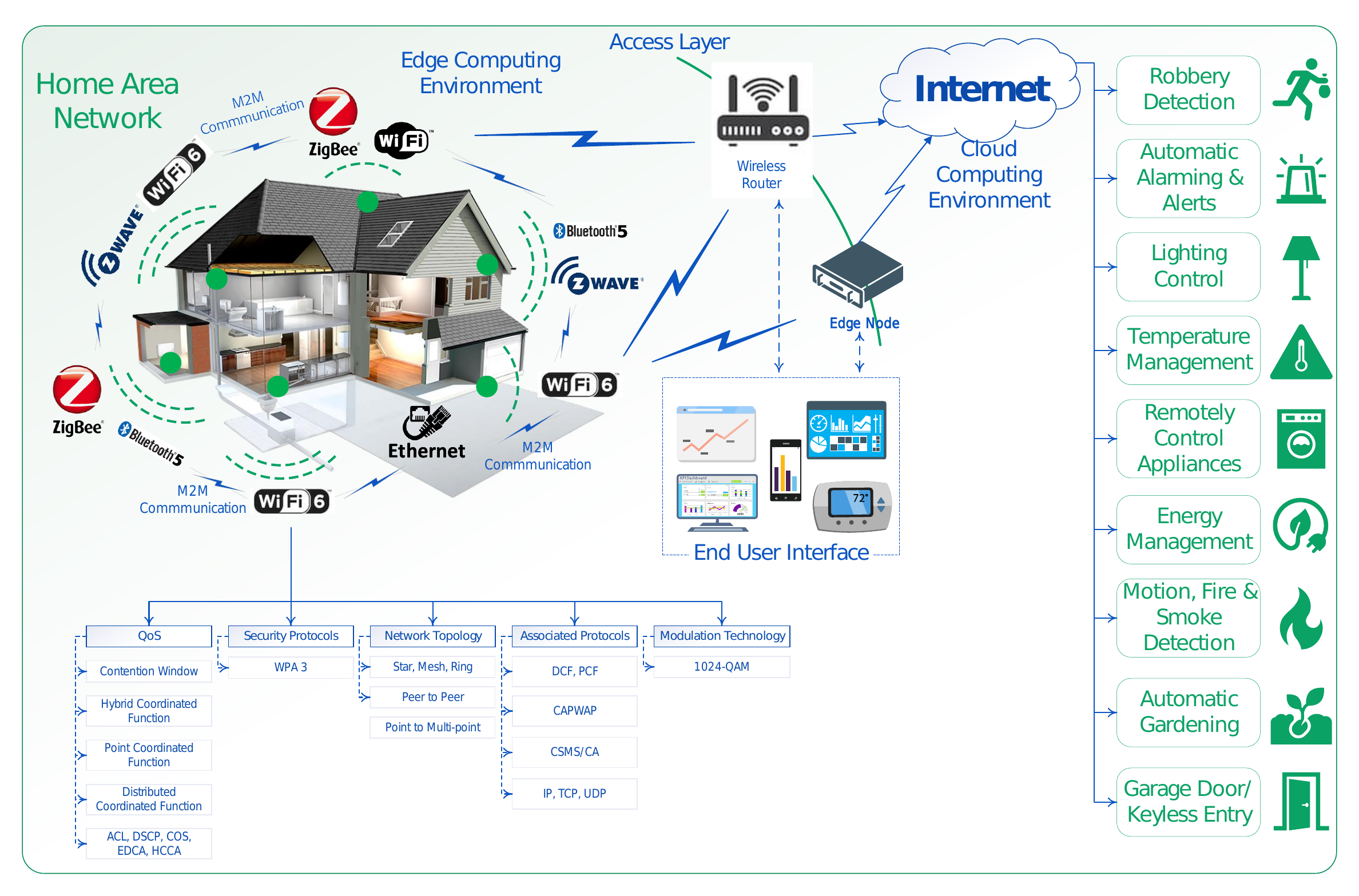}
\caption{Communication scenario for intelligent home applications.}
\label{fig_sim4}
\end{figure*}

As an example of a medium-range intelligent home application, Fig. \ref{fig_sim4} illustrates the communication technologies deployed in a HAN combined with the Analytics Everywhere framework to support automated analytical tasks. 
In intelligent homes, as in other IoT use cases, seamless connectivity is an important factor that gives users the flexibility to govern smart sensors from any location \cite{mao2023review}. Advanced communication tactics can minimize a smart device's power consumption and augment its communication efficiency \cite{sovacool2020smart}. Wi-Fi 6 and Wi-Fi 6E provide 1000Mbps of data rate with almost 1.5ms of negligible latency. Moreover, it uses a 6GHz band, which gives less interference compared to other technologies and provides excellent QoS with the help of mechanism like hybrid coordination function (HCF) which includes enhanced distributed channel access (for prioritized QoS services), and HCF controlled access (for parameterized QoS services). 

Federal Communication Commission (FCC) recently adopted a 6GHz band with new rules for unlicensed spectrum use. 6GHz spectrum has already been used by some licensed services (point-to-point microwave links, fixed satellite systems, etc.), which could increase the chance of harmful interference with unlicensed services (specifically outdoor operations). FCC permits unlicensed devices to operate at very low-power for indoor operations, and at standard-power with automated frequency control (AFC) mechanism for outdoor operations across the 6GHz band in order to avoid collision with licensed services. 

The key advantages of Wi-Fi 6 for IoT are TWT, dual carrier modulation (DCM), and the ability to offer lower bandwidth to IoT sensors (a single Resource Unit of 2MHz can be offered to provide 375 Kbps, which is ideal for IoT sensors). This effectively improves the link budget by 8 dB hence improving the range. Wi-Fi 6 is based on orthogonal frequency-division multiple access (OFDMA), allowing simultaneous sessions to transmit together using resource units and trigger frames.  Considering all these facts, Wi-Fi 6 can potentially help in reaching the final goal of intelligent homes, which is to increase comfort and quality of life for intelligent home residents. In a HAN, accumulated and continuous IoT data streams from smart devices can normally be analyzed at the far edge and cloud using an Analytics Everywhere framework \cite{cao2019analytics}. The insights from this process can support robbery detection, energy optimization, and automated gardening.

\subsection{Transportation/Mobility Sector}

 Most significant IoT applications in this sector are navigation and route optimisation, including real-time fleet monitoring, transit operation optimisation, parking availability monitoring, and driver behaviour tracking. Table \ref{tab:WAN1} presents different long-range communication technologies that provide MAN/WAN/LPWAN networking. Cellular, Cat-M, Cat-0, and 5G are the principal WAN networks, while Weightless W/N/P, LoRaWAN, SigFox, NB-IoT, and EC-GSM-IoT are LPWAN communication technologies. Most LPWAN network types serve QoS except LoRaWAN, which is based on slotted alohanet. Most of the LPWAN networks are equipped with better security mechanisms.  NB-IoT is the preference for QoS among LPWAN \cite{kanj2020tutorial}.
 
 LoRaWAN follows a separate network architecture and, depending on the sensor node, requires a gateway to communicate with another endpoint, which usually is a star network topology \cite{miles2020study}. LoRaWAN supports 20-65km (dependent on many conditions e.g. spreading factor) of coverage range in a rural area with 13km of stable distance with a 500ms of latency. In contrast, SigFox has a higher latency of 2s in almost 50km of rural coverage area and 11km of stable range. In terms of data rate, LoRaWAN outperforms SigFox. In recent years, Weightless W has been one of the rarely used LPWAN communication technologies, though it has some advantages like maximum data rate and minimum power ratio among LPWAN protocols. However, LoRaWAN and SigFox seem to own the market in unlicensed spectrum. Therefore, the choice of communication technology relies on the type of IoT application being used.

\begin{table*}[ht!]
\centering
\caption{Communication Technologies Suited for IoV Applications}
\label{tab:WAN1}
\resizebox{\linewidth}{!}{
\begin{tabular}{llllllllll}
\toprule
\begin{tabular}[c]{@{}l@{}}\textbf{Communication} \\ \textbf{Technologies}\end{tabular} & \begin{tabular}[c]{@{}l@{}}\textbf{IPv6}\\ \textbf{Support}\end{tabular} & \textbf{Duplex Mode} & \textbf{Type of Network} & \textbf{QoS} & \textbf{Latency (ms)} & \textbf{Data Rate (Mbps)} & \begin{tabular}[c]{@{}l@{}}\textbf{Power}\\ \textbf{Ratio}\\\textbf{(dBm)}\end{tabular} & \begin{tabular}[c]{@{}l@{}}\textbf{Power} \\ \textbf{Saving} \\ \textbf{Mechanism}\end{tabular} & \textbf{Coverage Range (m)} \\
\midrule
Cellular & Yes & Full & MAN / WAN & Yes & 10, 50 & \begin{tabular}[c]{@{}l@{}}1G:0.002, 2G:0.064, \\3G:0.144, 4G:100 - 1000  \end{tabular}& 27, 37 & Yes & 8000 - 40000 \\
Cat-M & No & Half / Full & MAN / WAN & Yes & 10 - 15 & 1 & 23 & Yes & 5000 \\
Cat-0 & No & Half / Full & MAN / WAN & Yes & 18 & 1 & 23 & Yes & 5000 \\
DigiMesh & Yes & Half / Full & WPAN & No & 13 & 0.1152 & 10 - 18 & Yes & 90 - 1600 \\
EC-GSM-IOT & Yes & Half / Full & LPWAN & Yes & 700 - 2000 & 0.24, 0.35 & 23 - 33 & Yes & 15000 \\
LoRaWAN & Yes & Half / Full & LPWAN & Yes & 500 & 0.05, 0.027 & 20 & Yes & 5000, 13000, 20000 \\
NB-IOT & Yes & Half & LPWAN & Yes & $<$10000 & 0.2, 0.234, 0.2048 & 23 & Yes & 100 - 10000 \\
Optical Fiber & Yes & Half / Full & LAN / MAN / WAN & Yes & 0.005 & 100 - 43000000 & 30 - 37 & Yes & 550, 1000, 2000, 10000 \\
RF & Yes & Half / Full & N/A & Yes & 4.2 & 0.02 - 10000 & 20 - 40 & No & 0.0001 - 100000000 \\
SigFox & Yes & Half & LPWAN & Yes & 2000 & 0.0001, 0.0006 & 21.7 & Yes & 10000, 11000, 50000 \\
WirepasMesh & Yes & Half / Full & WPAN & Yes & 10 & 1 & 5 & No & 100 - 10000 \\
WiMAX & Yes & Full & WMAN & Yes & 40 - 60 & 30, 75, 100 & 23 - 43 & Yes & 3500, 10000 \\
Wi-Fi HLow & No & Half & CAN / WLAN & Yes & 64 & 0.15 – 4, 0.65 - 7.8 & $<$30 & Yes & 100 – 1000, 700 \\
Weightless N & No & Full & LPWAN & No & 8000 - 12000 & 0.03 - 0.1 & 17 & No & 2000 \\
Weightless W & No & Half & LPWAN & Yes & 8000 - 12000 & 0.001 - 10 & 17 & No & 5000 \\
Weightless P & No & Full & LPWAN & Yes & 8000 - 12000 & 0.0002 – 0.1 & 17 & No & 2000 \\
5G & Yes & Full & MAN / WAN & Yes & 1 - 10 & $>$=1000 & 33 - 43 & Yes & 20000, 45000, 60000 \\\bottomrule
\end{tabular}}
\end{table*}

\begin{figure*}[!ht]
\centering
\includegraphics[width=\linewidth]{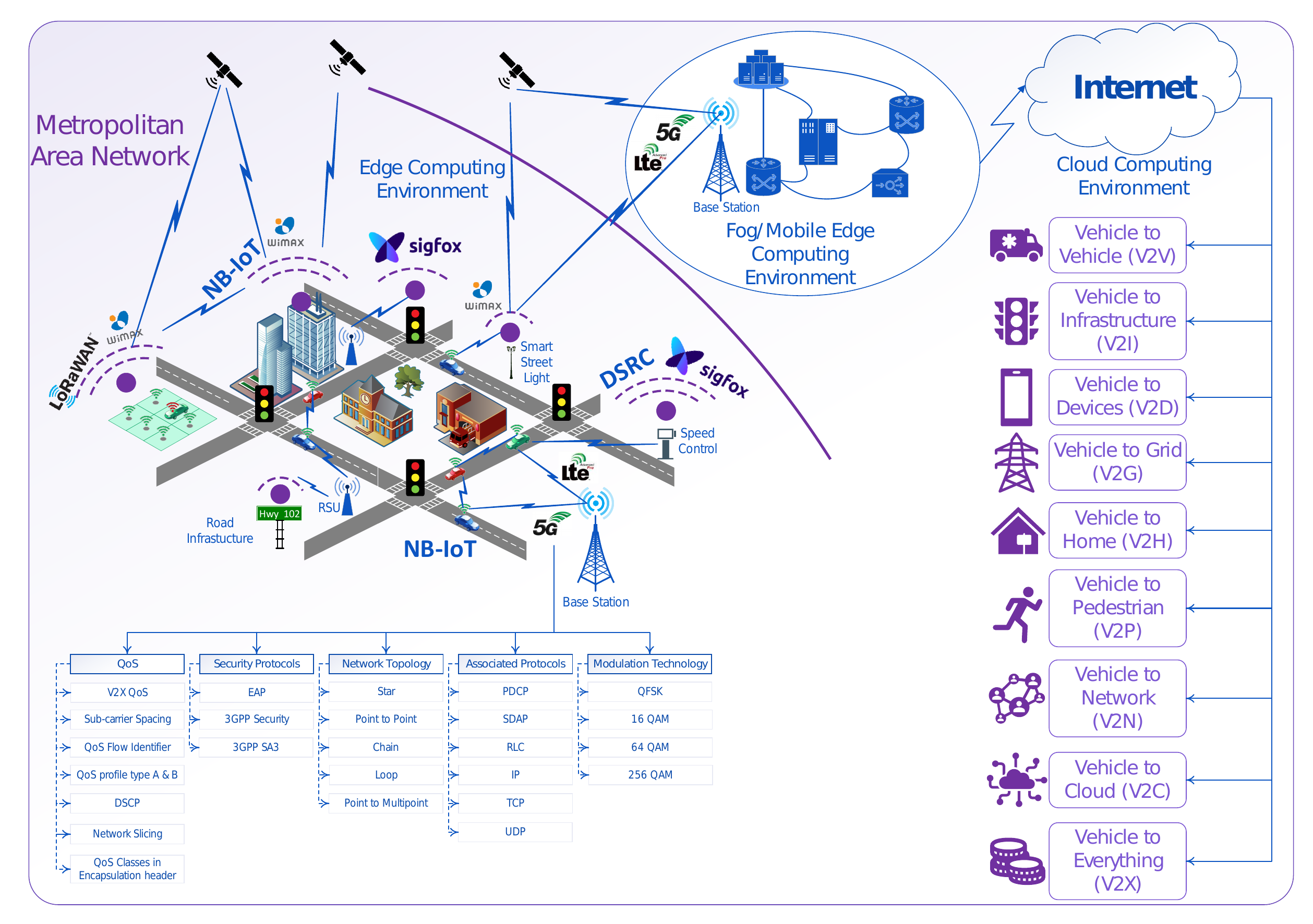}
\caption{Communication scenario for IoV applications.}
\label{fig_sim5}
\end{figure*}

5G paves the way for IoT with the highest connection density of 1 million devices/km$^2$ compared to other communication technologies \cite{van20225g}. That such an immense number of devices could be able to communicate with the Internet within a given space is a gigantic step forward for an application like IoV. Fig. \ref{fig_sim5} illustrates the networking within the context of IoV, where far-edge/near-edge/cloud computing continuum can support complex automated ML tasks. 
In an IoV application, 5G technology plays a pivotal role since it provides a special QoS for IoV called V2X QoS. It also provides more than 1000Mbps data rate and supports vehicle telematics, autonomous vehicles, and in-vehicle infotainment applications. In Cellular V2X (C-V2X), there are two interfaces: Uu (long-range cellular network communication) and PC5 (short-range, network-less, direct communication). The Uu utilizes the 5G spectrum while PC5 is short-range, similar to what was supported with DSRC (Dedicated short-range communications). V2V, V2I, and V2P communications operate in ITS bands (e.g. 5.9GHz), which are independent of the cellular network, whereas V2N communication operates in mobile broadband licensed spectrums such as V2V, V2I, V2H, and V2X.

Transportation is, in fact, a unique vertical sector because it is already adopting new universal architectures that rely on different communication networks to serve the needs of smart cities \cite{duan2020emerging}. However, the increasing number of moving smart vehicles with different speeds is already triggering problems in packet delivery to sink nodes, as well as packet delivery rate. This causes a ripple effect on the gathering of continuous data streams, the performance of ML tasks, and the development of future V2X applications.


Many ML algorithms have been proposed for smart transportation applications, including AdaBoost, Bayesian Network Seasonal Autoregressive Integrated Moving Average (BN-SARIMA), Coupled Hidden Markov Model (CHMM), Convolutional Neural Network (CNN) and Deep CNN (DCNN), and Decision Trees. Please refer to \cite{zantalis2019review} for an extensive review. With the Intelligence Everywhere learning paradigm, some of the aforementioned ML algorithms can easily be adapted to fit in federated learning and incremental learning mechanisms to provide better smart transportation applications. However, it is important to point out that the potential of V2X in developing new IoT applications serving the future needs of smart cities has not yet been fully investigated.  Other network performance metrics (e.g. throughput, end-to-end delay, latency) also need to be investigated for developing innovative real-time IoV applications, taking into account the privacy and security attacks.


\subsection{A Vision for Future Horizontal IoT applications}
IoT platforms have been developed globally to support different applications ranging from industrial/manufacturing and transportation/mobility to energy, healthcare, and supply chains. The current landscapes of IoT platforms highlight that most of them are focusing on specific sectors, also known as verticals \cite{scully_2020}. 
The current trend in IoT is moving towards the adoption of horizontal applications that span across different sectors, promoting interoperability in diverse IoT networks. These horizontal IoT applications will play a key role in supporting complex use cases involving multiple sectors simultaneously.
For example, we might see a new IoT application that is synergy in the healthcare and smart building sectors or an IoT application embedded in transportation/mobility, energy, and supply chain sectors.

\begin{figure}[!ht]
\centering
\includegraphics[width=\linewidth]{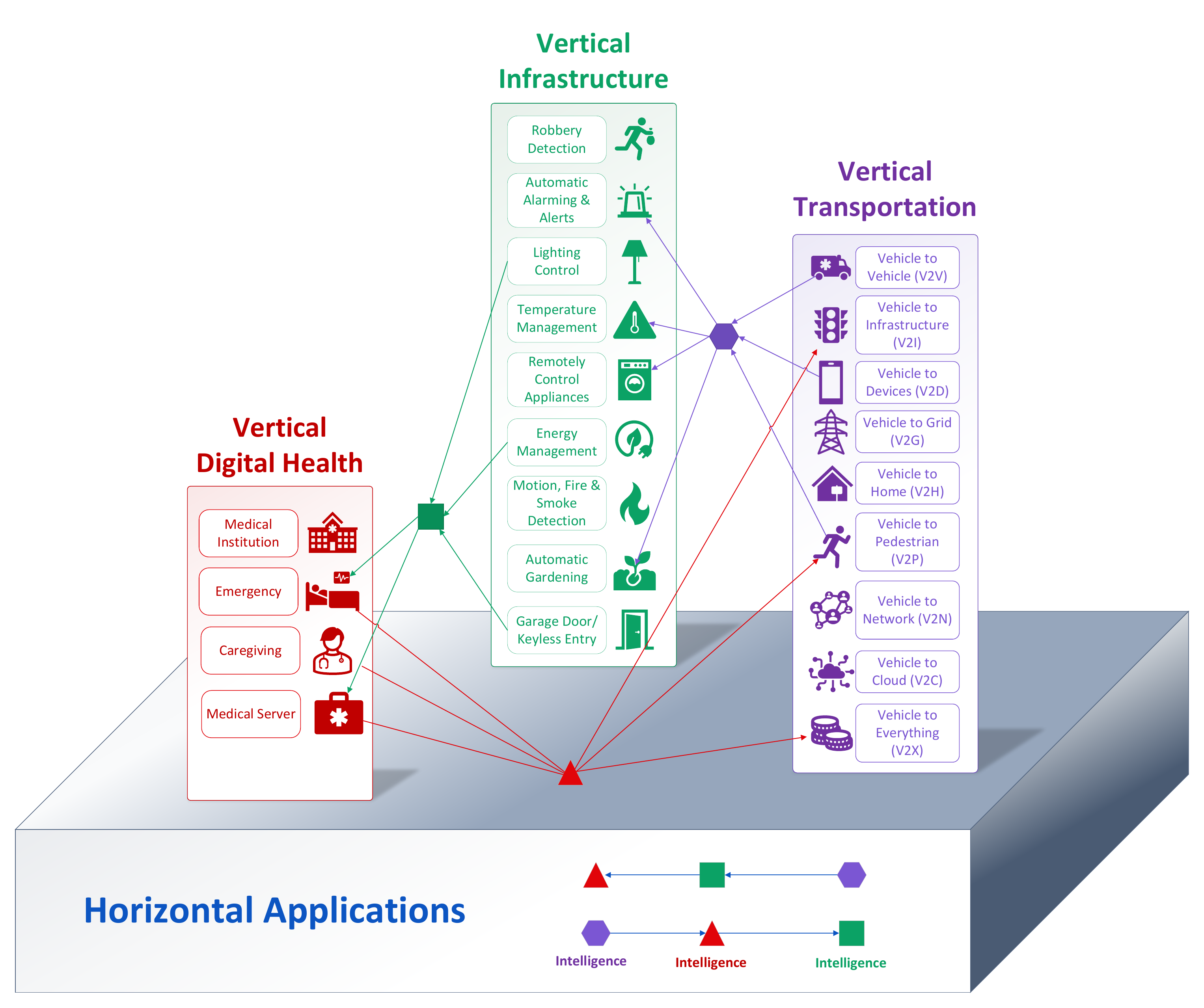}
\caption{Communication scenario for horizontal IoT applications.}
\label{fig-horizontal-app}
\end{figure}

As we are moving towards a Society 5.0 in the very near future \cite{fukuda2020science}, horizontal IoT applications will play a core role in transforming a large amount of streamed data across multiple vertical sectors into collective intelligence that can be made transferable, inter-connectable, and
collectively create more value. Figure \ref{fig-horizontal-app} depicts an example of a horizontal IoT application that crossed multi-vertical sectors, including digital health, infrastructure, and transportation sectors. 

We envisage that the collective intelligence obtained from the analytical activities or ML models in a vertical IoT application (i.e., digital health) can be re-used and transferred to other vertical IoT applications (i.e., infrastructure or transportation) as input information for other analytical activities or other ML models. For example, through the concept of collective intelligence, health analytics information of the home/building users is learned at the edge and transmitted to the cloud, where it is aggregated and processed to optimize the home/building environment, leading to improved health outcomes for the users. 

Simultaneously, the knowledge gained from the smart home/building sector is fed back to the users at the edge, enhancing their awareness of other issues, such as energy optimization and behavior changes, fostering a more sustainable and energy-efficient home/building ecosystem. The collective intelligence system continually learns from data gathered across multiple homes/buildings and user interactions. This ongoing learning process enables the system to improve its recommendations and responses over time, enhancing the overall effectiveness and intelligence of the IoT networks.


From the Intelligence Everywhere perspective, data, and insights can traverse through multiple vertical applications using a well-defined data lifecycle and the IoT networks outlined in Tables \ref{tab:BAN}, \ref{tab:HAN}, \ref{tab:WAN1}. Learning capability and analytical capability can assist developers in providing a mix of applications in which analytical results from the digital health sector or transportation can be used/transferred as the input for the smart home/building and vice versa.

%% file: CONTENTS/6-Challenges.tex
\section{Challenges and Opportunities}

In this section, we have identified the open challenges and opportunities to achieving an Intelligence Everywhere learning paradigm. This list is not meant to be exhaustive, but it gives an indication of topics for future research and development.

\subsection{Challenges}

A large number of important questions remain open on the topics of security, privacy, scalability, compatibility, reliability, and resilience. They are explained in more detail below.

\begin{itemize}
    \item   \textit{Security}: Most mobile IoT devices are tied through wireless technologies with internal APs, switches,  routers, and near-edge nodes. If the end IoT device/node/sensor is compromised, the door is opened for a collapse of the entire organisation’s network or a malicious interference on the data life-cycle of an Intelligence Everywhere learning paradigm, especially in the case of descriptive analytics tasks that might be running at the far-edge of a network. There are actually more chances of success in exploiting breaches with the IoT devices compared to the core networking devices (AP, Switch, Router, Firewall), mainly because IoT sensors are fully exposed to the outside world. NISTIR 8259 illustrates the recommended cybersecurity activities for manufacturers that can help them make IoT devices more securable. IETF open standard manufacturer usage description (MUD - RFC 8520) provides features like device visibility and context-specific access policies, which reduce the threat surface for IoT devices. Intelligence Everywhere learning paradigms will play an important role in developing new approaches for IoT security mainly because the next generation of far-edge/near-edge/cloud ML tasks will be running in real-time. However, it will expose new security challenges in the domain of collective intelligence once human and IoT systems start to collaborate in the near future.
    \item  \textit{Privacy}: IoT devices usually cultivate, fetch, and transmit sensitive data streams. These data could be personal, private, or business-critical. Thus, maintaining confidentiality at all levels during transportation and manipulation is a pertinent task. This is imperative in the context of collective intelligence. With the next generation of encryption algorithms, we are moderately safe, but IoT will require new privacy-aware algorithms at the device level, analytical capability level, and user level. NISTIR 8228 gives great insight into cybersecurity and privacy risk mitigation for the IoT environment. Enforcement of GDPR (General Data Protection Regulation) in IoT has a very large and encompassing set of requirements like personally identifiable information, privacy, and data protection by design for the entire fabric of IoT. Fulfilling such prerequisites enhances privacy in IoT. We would like to point out that only when the IoT industry overcomes these privacy issues, will it be able to earn the trust level required to bring new business models to horizontal IoT applications.
    \item  \textit{Scalability}: With the deployment of more advanced IoT devices in the near future, the interconnected network will also be expanded. Therefore, the design of an IoT network architecture requires scalability, which still remains a technical challenge. If an IoT network is not scalable, an Intelligence Everywhere learning paradigm is also not scalable. Also, it remains unclear how collective intelligence will impact the scalability of IoT applications. We will only be able to serve very limited devices and run very simple automated ML tasks under a single infrastructure. This creates a whole new independent environment for new IoT devices. A scalable IoT network is a cost-effective solution that conveniently ratifies new smart IoT devices within the same network.
    \item  \textit{Compatibility}: Another challenge is compatibility, which is related to better communication and analytical accuracy. An old version of the firmware/software of the IoT sensor or an old version of an ML algorithm needs to be compatible with new versions, as some vendors provide over-the-air firmware/software updates. If the mismatch of firmware versions does not let two devices talk, this will lead to a communication breakdown that can cause problems in a network and for analytics. Hence, the importance of addressing new compatibility solutions in IoT networks.
    \item  \textit{Reliability}: When two devices communicate with each other in an IoT use case, there will always be a chance of data loss due to certain parameters, like collisions, that can curtail reliability. Research on reliability could be expanded by a finer choice of communication technologies that have a better mechanism for error handling and re-transmission of lost information [8]. From the Intelligence Everywhere learning perspective, process mining becomes imperative to simultaneously monitor both the computing resources and data life-cycles.
    \item  \textit{Resilience}: Always connected things in IoT is one of the crucial requirements since failures and connection loss can cause fatal accidents. Due to different fault types (physical, interaction, transient, permanent, and development), it is challenging to detect anomalies and improve the resilience of IoT networks \cite{ratasich2019roadmap}. 
    
\end{itemize}

\subsection{Opportunities}

At the same time, new opportunities emerge in the fields of infrastructure supervision and context awareness.

\begin{itemize}
    \item \textit{Human-Machine Collaboration for Exploitation-Exploration:} Intelligence Everywhere is promising to evolve to become an interconnected and collective intelligent ecosystem where devices, systems, services, and users work together to support IoT applications. Normally, it requires some tradeoff between the utilization of known solutions (exploitation) and new solutions (exploration) to achieve the highest level of efficiency for collective intelligence.  The synergy of artificial/machine intelligence and human intelligence are usually seen as complementary forms of intelligence \cite{casadei2023artificial}. Therefore, Intelligence Everywhere with a variety of IoT networks, and multi-models of learning mechanisms in combination with human input will potentially open many opportunities and directions for looking for new ways to achieve common goals such as using the wisdom of crowds, using multi-agent systems, markets, swarm intelligence. 

    \item  \textit{Infrastructure Supervision}: Supervising millions of endpoints should be of top priority for IoT networks because, once all the nodes are deployed and are transferring new insights, we need to know whether they are all still up and running properly and efficiently. In infrastructure supervision, both device monitoring and network monitoring are covered. IoT network observation is an equally compelling issue to address as the communication process of the IoT relies on underlying networks. Infrastructure supervision can slash downtime and help engineers to quickly troubleshoot problems. It will be a win-win situation for any IoT application if we have end-to-end device and network visibility.
    
    \item  \textit{Context Awareness}: After the connection is entrenched between IoT devices, the correct choice of communication technologies could make data transfer very nimble, with limited latency. Intelligence Everywhere Learning supports on-premise and real-time data processing as needed with mobility features. With the help of context sensing, collective intelligence, and incremental/federated learning, it opens new doors for ML tasks like deep visibility. Therefore, the far-edge/near-edge/cloud continuum can enhance the routes of automation in IoT networking, and provide visibility to the end user/application. As well, with device comparison, it improves the performance and security of IoT devices and network elements. Collective Intelligence is being inserted in all aspects of the end-to-end IoT ecosystem, from the device all the way to the cloud. Finally, merging IoT networking and ML models can lead to enormous opportunities for IoT research and development to support the continued scaling and build-out of IoT into nearly everything.

\end{itemize}

%% file: CONTENTS/7-Conclusion.tex
\section{Conclusions}


In the fast-paced development of the Internet of Things, the importance of collective intelligence becomes evident as communication networks evolve from their traditional role of connecting "things" to becoming vehicles for creating and transferring valuable "insights." Embracing this vision, this article introduces the Intelligence Everywhere learning paradigm, emphasizing its four core capabilities: resource, analytical, learning, and data life-cycle capabilities. By focusing on these capabilities, the aim is to assist developers in building new platforms that not only enhance existing vertical IoT applications but also envision and pave the way for pioneering horizontal IoT applications. The incorporation of collective intelligence through the Intelligence Everywhere paradigm empowers IoT networks to harness the collaborative potential of diverse devices, systems, and users, ultimately leading to more intelligent and innovative IoT applications that address complex challenges and foster seamless interoperability across multiple domains.

We have identified the state-of-the-art in
IoT networking and machine learning from a data scientist perspective, which has culminated into a collection of open challenges and opportunities. However, it is important to point out that achieving an Intelligence Everywhere learning paradigm can only be successful if users actually feel safe in an intelligent society, and know exactly what is happening with their data.